\documentstyle[aps,prl,floats]{revtex}

\newcommand{\bastar}{\begin{eqnarray*}}
\newcommand{\eastar}{\end{eqnarray*}}
\newskip\humongous \humongous=0pt plus 1000pt minus 1000pt

\newif\ifdtup

\relax
\newcommand{\be}{\begin{equation}}
\newcommand{\ee}{\end{equation}}
\newcommand{\bea}{\begin{eqnarray}}
\newcommand{\eea}{\end{eqnarray}}
\newcommand{\X}{{\vec X}}
\newcommand{\pro}{\partial}
\newcommand{\n}{\hat n}
\newcommand{\oneg}{\displaystyle\frac{1}{g}}

\newcommand{\D}{{\hat D}}

\newcommand{\A}{{\vec A}}
\newcommand{\valpha}{{\vec \alpha}}

\newcommand{\dfrac}{\displaystyle\frac}
\newcommand{\ba}{\begin{array}}
\newcommand{\ea}{\end{array}}

\newcommand{\nn}{\nonumber}
\newcommand{\hn}{\hat n}
\begin{document}
\twocolumn[\hsize\textwidth\columnwidth\hsize\csname@twocolumnfalse%
\endcsname
\title  {Monopoles and Knots in Skyrme Theory}
\bigskip

\author{Y. M. Cho}

\address{
Department of Physics, College of Natural Sciences, Seoul National University,
Seoul 151-742, Korea  \\
{\scriptsize \bf ymcho@yongmin.snu.ac.kr
} \\ \vskip 0.3cm
}
\maketitle

\begin{abstract}
~~~~~We show that the Skyrme theory actually is a theory of monopoles
which allows a new type
of solitons, the topological knots made of monopole-anti-monopole pair,
which is different from the well-known skyrmions. Furthermore
we derive a generalized Skyrme action from
the Yang-Mills action of QCD, which we propose to be
an effective action of QCD in the infra-red limit.
We discuss the physical implications of our results.

\vspace{0.3cm}
\end{abstract}

\narrowtext
\bigskip
                           ]
The proposal that the Skyrme theory could describe
an effective theory of QCD in the low energy limit has become very
popular \cite{s1,brown}. In this view the skyrmions are
interpreted as the baryons, and numerous
evidences have been put forward which support this view \cite{jack}.
Motivated by the success of the proposal many people
have tried to derive the Skyrme action from QCD.
But a concrete theoretical proof
of the proposal, that one can actually derive
the Skyrme action from QCD, has remained very difficult \cite{simic}.

The purpose of this Letter is two-fold. First we show that the Skyrme
theory actually has a much richer soliton spectrum. Indeed
we show that it is a theory of self-interacting non-Abelian monopoles, and
prove that it allows
a new type of knotted solitons very similar to
the topological knots in the Skyrme-Faddeev theory of
non-linear sigma model \cite{f1}.
The other purpose is to present a new
evidence that the Skyrme theory is indeed very closely related
to QCD. In fact, by reparametrizing the gauge potential,
we derive a generalized Skyrme action from
the Yang-Mills action of QCD which we propose to be 
an effective theory of QCD in the low energy limit.
The assertion that the Skyrme theory could be derived from
QCD is perhaps not suprising. But the fact that the Skyrme
theory has a new type of solitons is really unexpected,
which put the Skyrme theory in a completely new perspective.

Let us start from the Skyrme theory. With
\bea
&U = \exp (\xi \dfrac{\vec \sigma}{2i} \cdot \hat n)
= \cos \dfrac{\xi}{2} - i (\vec \sigma \cdot \hat n) \sin \dfrac{\xi}{2}, \nn\\
&L_\mu = U\partial_\mu U^{\dagger}, ~~~~~({\hat n}^2 = 1)
\eea
the Skyrme Lagrangian is expressed as
\bea
&{\cal L} = \dfrac{\mu^2}{4} {\rm tr} ~L_\mu^2 + \dfrac{\alpha}{32}
{\rm tr} \left( \left[ L_\mu, L_\nu \right] \right)^2  \nn \\
&= - \dfrac{\mu^2}{4} \Big[ \dfrac{1}{2} (\partial_\mu \xi)^2
+(1-\cos\xi)(\partial_\mu \hat n)^2 \Big]  \nn \\
& -\dfrac{\alpha}{16} \Big[ \dfrac{1-\cos\xi}{2} (\partial_\mu
\xi \partial_\nu \hat n - \partial_\nu \xi \partial_\mu \hat n)^2  \nn \\
& + (1-\cos\xi)^2 (\partial_\mu \hat n \times \partial_\nu \hat
n)^2 \Big].
\eea
The equation of motion is given by
\bea
&\dfrac{\mu^2}{4} \left[ \partial^2 \xi -\sin\xi (\partial_\mu \hat
n)^2 \right] \nn\\
&+\dfrac{\alpha}{32} \sin\xi (\partial_\mu
\xi \partial_\nu \hat n  -\partial_\nu \xi \partial_\mu \hat n)^2 \nn\\
&+\dfrac{\alpha}{8} (1-\cos\xi) 
\partial_\mu \big[ (\partial_\mu
\xi \partial_\nu \hat n  -\partial_\nu \xi \partial_\mu \hat
n) \cdot \partial_\nu \hat n \big] \nn\\
& - \dfrac{\alpha}{8} (1-\cos\xi) \sin\xi (\partial_\mu \hat n \times
\partial_\nu \hat n)^2 =0, \nn \\
&\partial_\mu \Big\{ \dfrac{\mu^2}{4} (1-\cos\xi) \hat n \times
\partial_\mu \hat n \nn\\
& + \dfrac{\alpha}{16} (1-\cos\xi) 
\big[ (\partial_\nu \xi)^2
\hat n \times \partial_\mu \hat n
-(\partial_\mu \xi \partial_\nu \xi) \hat n \times
\partial_\nu \hat n \big] \nn\\
&+\dfrac{\alpha}{8} (1-\cos\xi)^2
(\hat n \cdot \partial_\mu \hat n \times
\partial_\nu \hat n) \partial_\nu \hat n \Big\}=0.
\eea
With the spherically symmetric ansatz
\bea
\xi = \xi (r),~~~~~\hat n = \hat r,
\eea
(3) is reduced to
\bea
&\dfrac{d^2 \xi}{dr^2} +\dfrac{2}{r} \dfrac{d\xi}{dr} -\dfrac{2
\sin\xi}{r^2}
-\dfrac{\alpha}{\mu^2} \Big[ \dfrac{\sin^2 (\xi /2)}{r^2}
\dfrac{d^2 \xi}{dr^2} \nn\\
&+\dfrac{\sin\xi}{2 r^2} (\dfrac{d\xi}{dr})^2
-\dfrac{2 \sin\xi \sin^2 (\xi /2)}{r^4} \Big] =0.
\eea
Imposing the boundary condition $\xi(0)= 2\pi$ and $\xi(\infty)= 0$,
one has the well-known skyrmions. The reason for the solitions, of
course, is the non-trivial homopopy $\pi_3 (S^3)$ defined by (1)
\cite{s1,brown}.

Now, we show that (3) actually allows a new type of knot solitons.
To see this notice that with
\bea
\xi = \pi,
\eea
(3) is reduced to the following equation for
$\hat n$
\bea
\hn \times
\partial^2 \hn + \dfrac{\alpha}{\mu^2} ( \partial_\mu H_{\mu\nu} )
\partial_\nu \hn = 0,
\eea
where
\bea
H_{\mu\nu} = \hn \cdot (\partial_\mu \hn \times \partial_\nu \hn)
= \partial_\mu C_\nu - \partial_\nu C_\mu. \nn
\eea
Notice that since $H_{\mu\nu}$ is closed, one can always 
introduce the potential $C_\mu$
for the field $H_{\mu\nu}$, as far as $\hn$ is smooth
everywhere.
But (7) is precisely the knot equation that we obtain from the Skyrme-
Faddeev action \cite{cho3}
\bea
{\cal L}_{SF} = - \dfrac{\mu^2}{2} (\partial_\mu \hat n)^2 -
\dfrac{\alpha}{4} (\partial_\mu \hat n \times \partial_\nu \hat n)^2.
\eea
This immediately tells that the knots in this theory automatically become
the solutions of the Skyrme theory. This guarantees the existence
of the knots in Skyrme theory. This is striking. But considering the fact that
the Skyrme action (1) is indeed an obvious
generalization of the Skyrme-Faddeev action (8), one should
not be too suprised at the knots.

Another remarkable point is that the Skyrme theory actually
has the non-Abelian monopoles as the classical solution.
This must be clear because the knot equation (7) allows
the monopoles \cite{cho3,cho80}. Indeed it is clear that
\bea
\xi = \pi,~~~~~\hat n = \hat r,
\eea
forms a solution of (3) except at the origin, because
\bea
\partial^2 \hat r = - \dfrac {2}{r^2} \hat r, ~~~\partial_\mu H_{\mu\nu} =0.
\eea
Notice that (9) forms a solution of (3) even without the
non-linear interaction (i.e., with $\alpha=0$). As we will see
soon the solution can naturally be identified as the monopole \cite{cho3,cho80}.

It is really remarkable that the Skyrme theory has such a rich
spectrum of classical solutions, all of which are of topological
origin. Skyrme noticed that the existence of
$\pi_3 (S^3)$ described by (1) produces the topological
skyrmions. But our analysis tells that the topological field
$\n$ defines two more homotopies $\pi_3 (S^2)$ and $\pi_2 (S^2)$,
which produce their own knots and monopoles.

At this point we may ask what is the relationship among
these topological solutions. To see this remember that
the baryonic charge of the skyrmions is given by
\bea
&Q_s = \dfrac{1}{24\pi^2} \int \epsilon_{ijk} ~{\rm tr} ~(L_i L_j L_k) \nn\\
&= \dfrac{1}{16\pi^2} \int \epsilon_{ijk} (1- \cos \xi) H_{ij}
\partial_k \xi d^3x.
\eea
On the other hand the knot and monopole charges are given by \cite{cho3,cho80}
\bea
&Q_k = \dfrac{1}{32\pi^2} \int \epsilon_{ijk} C_i H_{jk} d^3x, \nn\\
&Q_m = \dfrac{1}{8\pi} \int \epsilon_{ijk} H_{ij} d\sigma_k,
\eea
where $C_\mu$ is the potential of $H_{\mu\nu}$ defined by (7).
From this we deduce the followings. First the skyrmions carry
the monopole charge of the topological field $\hn$.
This suggests that the skyrmions are actually monopoles
whose energy is made finite by the non-linear interaction
with the scalar field $\xi$. More importantly, with
the Hopf fibring of $S^3 \simeq S^2\times S^1$ of the $SU(2)$ space, $\xi$ adds
the $U(1)$ charge of the fibre $S^1$ to the monopoles,
which makes the baryonic charge different from
the monopole charge. Secondly the knots carry neither
the baryonic nor monopole charges (At the same time
the skyrmions and the monopoles have no knot charge,
because the potential of their magnetic field become singular).
This suggests that the knots should be interpreted as the ``mesons'',
or more precisely the ``glueballs'', made of 
the monopole-anti-monopole pairs. The non-linear
self-interaction of $\hn$ not only makes their energy finite,
but also provides the topological stability to the knots.

This picture helps us to clarify the physical content of
Skyrme theory. Skyrme suggested that it is probably a theory
of non-linear sigma field, which allows the topological skyrmions.
But our analysis suggests a different
interpretation. {\it The Skyrme theory should really be viewed
as a theory of the monopole field
$\hn$ interacting with the scalar field $\xi$.
The non-linear self-interaction of $\hn$ produces the knots
made of monopole-anti-monopole pairs, and $\xi$
adds the $U(1)$ charge of the Hopf fibre to the monopoles to
produce the topological skyrmions}.

Now, we establish a deep connection between Skyrme action and QCD.
For this we need a
better understanding of QCD. A unique feature of the non-Abelian
gauge theory is its non-trivial topology, and it is essential
to take care of these topological characters in the dynamics
for us to understand QCD. For this purpose it is very important
to reparametrize the non-Abelian gauge potential in terms of
the restricted potential of the maximal Abelian
subgroup $H$ of the gauge group $G$ and the gauge covariant valence
potential of the remaining $G/H$ degrees
\cite{cho1,cho2}. To see this consider
$SU(2)$ for simplicity. A natural way to take care of
the topological degrees into the dynamics is to introduce
a topological field $\n$ which selects the color direction
at each space-time point, and to decompose the
non-Abelian connection into the restricted potential which makes $\n$
a covariant constant and the valence potential
which forms a covariant vector field \cite{cho1,cho2},
\bea
 & \vec{A}_\mu =A_\mu \n - \oneg \n\times\pro_\mu\n+\X_\mu\nonumber
         = \hat A_\mu + \X_\mu, \nn\\
 &  (A_\mu = \n\cdot \vec A_\mu, ~ \n^2 =1,~ \hat{n}\cdot\vec{X}_\mu=0).
\eea
Notice that the restricted potential $\hat A_\mu$ is
precisely the connection which
leaves $\n$ invariant under the parallel transport,
\bea
\D_\mu \n = \pro_\mu \n + g {\hat A}_\mu \times \n = 0.
\eea
Under the infinitesimal gauge transformation
\bea
\delta \n = - \vec \alpha \times \n  \,,\,\,\,\,
\delta \A_\mu = \oneg  D_\mu \vec \alpha,
\eea
one has \cite{cho1,cho2}
\bea
&&\delta A_\mu = \oneg \n \cdot \pro_\mu \valpha,\,\,\,\
\delta \hat A_\mu = \oneg \D_\mu \valpha  ,  \nn \\
&&\hspace{1.2cm}\delta \X_\mu = - \valpha \times \X_\mu  .
\eea
This shows that $\hat A_\mu$ by itself describes
an $SU(2)$ connection which
enjoys the full $SU(2)$ gauge degrees of freedom. Furthermore
the valence potential $\vec X_\mu$ transforms covariantly
under the gauge transformation.
More importantly the decomposition is gauge-independent.
Once the gauge covariant topological field $\hat n$ is given,
the decomposition follows automatically independent of the choice
of a gauge.

This decomposition, which has recently
become known as the ``Cho decomposition'' \cite{f2}
or the ``Cho-Faddeev-Niemi
decomposition'' \cite{lang}, was
introduced long time ago in an attempt to demonstrate
the monopole condensation in QCD \cite{cho1,cho2}.
But only recently the importance of the decomposition
in clarifying the non-Abelian dynamics
has become appreciated by many authors \cite{f2,lang}.
Indeed it is this decomposition which has played a crucial role
to establish the Abelian dominance
in the Wilson loops in QCD \cite{cho00}, and
the possible connection between the Skyrme-Faddeev action and the
effective action of QCD \cite{cho4}.

An important feature of $\hat{A}_\mu$ is that,
as an $SU(2)$ potential, it retains the full
topological characteristics of the original non-Abelian potential.
This is due to the topological field $\hat{n}$ in it.
Clearly the isolated singularities of $\hat{n}$ defines $\pi_2(S^2)$
which describes the non-Abelian monopoles. Indeed $\hat A_\mu$
with $A_\mu =0$ and $\hat n= \hat r$ describes precisely
the Wu-Yang monopole \cite{cho80,cho1}. So one can identify
the non-Abelian magnetic potential by
\bea
\vec C_\mu= -\frac{1}{g}\hat n \times \partial_\mu\hat n ,
\eea
which generates the magnetic field
\bea
\vec H_{\mu\nu}&=&\partial_\mu \vec C_\nu-\partial_\nu \vec C_\mu+ g \vec
C_\mu \times \vec C_\nu
=-\dfrac{1}{g} \partial_\mu\hat{n}\times\partial_\nu\hat{n} \nn\\
&=& -\dfrac{1}{g} (\partial_\mu C_\nu - \partial_\nu C_\mu) \hat n.
\eea
Furthermore, with the $S^3$
compactification of $R^3$, $\hat{n}$ characterizes the
Hopf invariant $\pi_3(S^2)$ which describes
the topologically distinct vacua \cite{cho3,cho79}.
So the restricted potential inherits all the essential topological
properties of the non-Abelian gauge theory.

The decomposition (13) leads us to establish the common ground
between QCD and Skyrme theory, and to clarify
the meaning of the Skyrme theory. To see this notice that
with (17) and (18) the Skyrme-Faddeev Lagrangian (8) can be expressed as
\bea
{\cal L}_{SF} = - \dfrac{\alpha}{4} \vec H_{\mu\nu}^2 - \dfrac{\mu^2}
{2} \vec C_\mu^2.
\eea
This allows us to interpret the solution (9) as the monopoles,
and to translate the vacuum number of QCD to the knot quantum
number of Skyme theory \cite{cho3,cho79}.

With the decomposition (13) we have
\bea
\vec{F}_{\mu\nu}&=&\hat F_{\mu \nu} + \D _\mu \X_\nu -
\D_\nu \X_\mu + g\X_\mu \times \X_\nu,
\eea
so that the Yang-Mills Lagrangian could be expressed as
\bea
&{\cal L} = -\dfrac{1}{4} \hat F^2_{\mu \nu} - \dfrac{\mu^2}{2} \hat A_\mu^2
-\dfrac{1}{4}(\D_\mu\X_\nu-\D_\nu\X_\mu)^2 \nn\\
&-\dfrac{g}{2} {\hat F}_{\mu\nu} \cdot (\X_\mu \times \X_\nu)
-\dfrac{g^2}{4}
(\X_\mu \times \X_\nu)^2,
\eea
where we have inserted a mass term for the restricted potential $\hat A_\mu$.
One may object the introduction of the mass term, because
it will certainly violate the gauge invariance of QCD.
But notice that, after the dynamical symmetry breaking with
the monopole condensation, one could assume (simple mindedly) that
the restricted potential acquires a mass \cite{cho4,cho5}. This would justify
the introduction of the mass term in the
infra-red limit of QCD.

With this preparation we now derive a generalized Skyrme theory
from the Lagrangian (21). To do this let
\bea
\X_\mu = f_1 \partial_\mu \n + f_2 \n \times \partial_\mu \n,
\eea
and find that the Lagrangian is expressed as
\bea
&{\cal L} = -\dfrac{1}{4} \big[ F_{\mu \nu} +(\phi^* \phi
-\dfrac{1}{g}) (\hn \cdot \partial_\mu \hat n \times \partial_\nu \hat n) 
\big] ^2 \nn\\
&-\dfrac{i}{2} (D_\mu \phi)^* (D_\nu \phi) (\hn \cdot \partial_\mu \hat n 
\times \partial_\nu \hat n) \nn \\
&-\dfrac{1}{4} \big| D_\mu \phi \partial_\nu \hat n -D_\nu \phi
\partial_\mu \hat n \big|^2
-\dfrac{\mu^2}{2} \big[A_\mu ^2 +\dfrac{1}{g^2} (\partial_\mu \hat
n)^2 \big],
\eea
where
\bea
\phi = f_1 + if_2,
~~~~~D_\mu \phi = (\partial_\mu + i g A_\mu) \phi. \nn
\eea
Furthermore, with $\partial_\mu \phi = 0$ and $A_\mu = \partial_\mu \xi$,
we have
\bea
&{\cal L} = -\dfrac{1}{4g^2}(1-g \phi^* \phi)^2 (\partial_\mu \hat n
\times \partial_\nu \hat n)^2 \nn \\
&-\dfrac{1}{4} \phi^* \phi (\partial_\mu \xi \partial_\nu \hat n
-\partial_\nu \xi \partial_\mu \hat n)^2 \nn \\
&-\dfrac{\mu^2}{2} \big[ (\partial_\mu \xi)^2 +\dfrac{1}{g^2}
(\partial_\mu \hat n)^2 \big].
\eea
At this point a striking similarity between the Skyrme Lagrangian
(1) and the QCD Lagrangian (23) becomes unmistakable. Of course (23) is
different from (1). But notice that it has a definite advantage
over (1), in the sence that it comes directly from QCD. With this
observation we can propose (23) as a generalized Skyrme Lagrangian which
could replace (1) as an effective theory of QCD in the low energy limit.

We close with the following comments. \\
1) The assertion that one can derive the Skyrme action from QCD is
understandable. Recently we have been able to derive
a generalized Skyrme-Faddeev action from the one-loop effective action
of QCD which we obtained with the monopole background \cite{cho4,cho5}.
This derivation was different from our derivation of the
generalized Skyrme action discussed here. Our discussion here
is based on a reparametrization of the potential
and does not appear to resort to an effective action of QCD
(except the insertion of the mass term of the restricted potential).
But clearly the two results are consistent with each other.
Indeed, as far as the generalized Skyrme theory could be viewed as
a generalized Skyrme-Faddeev theory, one could say that the derivation
of the generalized Skyrme-Faddeev action from the effective
action of QCD \cite{cho4,cho5} provides another evidence
which supports our derivation in this paper. \\
2) Our result suggests that the Skyrme theory is a theory of monopoles
interacting with the scalar field $\xi$. But here we like to argue
that it is also a theory of confinement which confines the monopoles.
To see this notice that in spite of
the existense of the mass scale $\mu$ in the theory, the theory admits
a long range monopole solutions. Furthermore, the mass scale
does not describe the mass of the elementary fields of
the theory. It describes the energy scale of the solitons.
The classical monopole solutions carry an infinite energy and thus
an infinite mass, while the monopole-anti-monopole bound states
carry finite energy. This suggests that the theory, just like
the Skyrme-Faddeev theory, is a theory
of confinement \cite{cho3}. What is remarkable is that the confinement
is manifest already at the classical level. In QCD one has to
produce a mass scale to prove the confinement. But here we already have
the mass scale at the classical level which we do not have to produce
from quantum correction. \\
3) We have proposed the generalized Skyrme action (23), 
in stead of the Skyrme action (1), as an effective action of QCD
in the low energy limit. In spite of the obvious differences 
they have a striking similarity.
One common aspect is that both admit the knot solitons \cite{cho6}, which
we have interpreted as the ``glueballs''.
We emphasize,
however, that they are the magnetic glueballs which carry the color
magnetic flux, while the glueballs in QCD (if they exist)
are supposed to
carry the color electric flux. This tells that our solitons
should describe very exotic states.
But the existence of such magnetic glueballs has already been 
predicted long time ago \cite{cho1,cho2}. Once the monopole condensation 
sets in, one should expect
the fluctuation of the condensed vacuum.  But obviously the fluctuation
modes have to be magnetic, which could be identified as 
the magnetic glueballs (A new feature here is that they 
have a topological stability. This could be an artifact of
the effective theory, not a genuine feature of QCD).
If so, the remaining task is to look for a convincing experimental evidence
of the magnetic glueball states in hadron spectrum \cite{cho1,cho2}.

The mass of the lightest knot in Skyrme theory can be estimated
to be about 1.42 times the mass of the nucleon. 
A detailed discussion will be presented 
elsewhere \cite{cho6}.

{\bf ACKNOWLEDGEMENT}

We thank Professor C. N. Yang for
the illuminating discussions. The work is supported in part by
a Korea Research Foundation Grant
(KRF-2000-015-BP0072), and by the BK21 project of the Ministry of Education.

\end{document}